\documentclass[aps,a4paper, preprint]{revtex4-1}
\usepackage{graphicx}
\usepackage{amsmath}
\usepackage{amssymb}
\usepackage{times}
\usepackage{dsfont}
\usepackage[T1]{fontenc}
\usepackage[utf8]{inputenc}
\usepackage{placeins}
\usepackage{makecell}
\usepackage{bm}
\usepackage[version-1-compatibility]{siunitx}
\usepackage[pdfborder=0]{hyperref}
\hypersetup{
    colorlinks=true,
    linkcolor=blue,          
    citecolor=blue           
}

\def\nn{\nonumber}
\def\beq{\begin{equation}}
\def\eeq{\end{equation}}
\def\beqna{\begin{eqnarray}}
\def\eeqna{\end{eqnarray}}
\def\bea{\begin{array}}
\def\ea{\end{array}}
\def\mg_+{{\mathcal A}}

\def\mg{{\mathcal G}}

\def\Tr{\mbox{Tr~}}
\def\det{\mbox{det\,}}
\def\etal{{\it et al.~}}
\def\Re{\operatorname{Re}}
\def\Im{\operatorname{Im}}
 
\begin{document}
\title{Deep noise squeezing in parametrically driven resonators}
\author{Adriano A. Batista$^1$}
\email{adriano@df.ufcg.edu.br}
\author{Raoni S. N. Moreira$^2$, and A. A. Lisboa de Souza$^3$}
\affiliation{
$^1$Departamento de Física, Universidade Federal de Campina Grande\\
Campina Grande-PB, CEP: 58109-970, Brazil\\
$^2$Centro de Ciências, Tecnologia e Saúde, Universidade Estadual da Paraíba,
Araruna-PB, CEP: 58233000, Brazil\\
$^3$Departamento de Engenharia Elétrica, Universidade Federal da Paraíba\\
João Pessoa-PB, CEP: 58.051-970, Brazil
}
\date{\today}
\begin{abstract}
Here we investigate white noise squeezing in the frequency domain
of classical parametrically-driven resonators with added noise.
We use Green's functions to analyse the response of resonators to added noise.
In one approach, we obtain the Green's function approximately using the
first-order averaging method,
while in the second approach, exactly, using Floquet theory.
We characterize the noise squeezing by calculating the statistical properties
of the real and imaginary parts of the Fourier transform of the resonators
response to added noise.
In a single parametric resonator, due to correlation, the squeezing limit of
$-6$~dB can be reached even with detuning at the instability threshold in a
single parametrically-driven resonator.
We also applied our techniques to investigate squeezing in a dynamical system
consisting a parametric resonator linearly coupled to a harmonic resonator.
In this system, we were able to observe deep squeezing at around $-40$~dB in
one of the quadratures of the harmonic resonator response.
We noticed that this occurs near a Hopf bifurcation to parametric instability,
which is only possible when the dynamics of the coupled resonators cannot be 
decomposed into normal modes.
Finally, we also showed that our analysis of squeezing based on Floquet theory
can be applied to multiple coupled resonators with parametric modulation and
multiple noise inputs.
\end{abstract}
\maketitle
\section{Introduction}
\label{intro}
Parametric resonators and amplifiers have been implemented in many different
systems of physics and engineering such as micro or nanoelectromecanical systems
(MEMS/NEMS)
\cite{karabalin2009parametric, thomas2013efficient,prakash2012parametric},
optomechanics \cite{he2023analytic}, Josephson Junctions
\cite{castellanos2008amplification,eom2012nature}, atomic force microscopy
\cite{moreno2006parametric}, ion traps \cite{paul90}, etc.  
Driving parametrically a resonator is a way to obtain very high effective
quality factors \cite{miller2018effective, lee2022giant}.
Hence, amplifiers based on these resonators can achieve very high gains and a
very narrow gain bandwidth.
The narrow band decreases the effect of fluctuations due to added noise.
Further decrease of fluctuations can be achieved when  noise squeezing
techniques are used.
These techniques have been used in high-precision position measurements 
\cite{szorkovsky2013strong} and also to measure small forces and helped detect
gravitational waves in the LIGO experiment
\cite{kimble2001conversion,chen2013macroscopic, abbott2016observation}.

In a pioneering paper in 1991, Rugar and Grütter \cite{rugar91} experimentally
observed thermal noise squeezing of flexural vibrations in a microcantilever. 
They used a two-phase lock-in amplifier to measure the effect, where the
signal was split into two channels in quadrature. The squeezing effect was most
noticeable near the transition line of the first parametric instability with the
parametric pump frequency set at twice the fundamental mode of the resonator.
They predicted  a lower bound of $-6$~dB amplitude deamplification at the
parametric instability threshold. 
They did not propose any theoretical model based on stochastic differential
equations to explain this effect. 

DiFilippo \etal\cite{difilippo92prl} developed noise squeezing
techniques for single ion mass spectroscopy based on
classical harmonic and anharmonic oscillators parametrically driven at twice
the natural frequency only.
Their model is based on Hamiltonian time
evolution of a Gaussian distribution of initial values, thus neglecting dissipation and noise.
They proposed two models of parametric squeezing which they called amplitude
(nonlinear squeezing)
and quadrature squeezing (around a saddle fixed point).
Their model cannot explain the sub-threshold thermomechanical
noise squeezing observed experimentally by
Rugar and Grütter.
In Ref.~\cite{cleland2005thermomechanical}, Cleland investigated thermal noise
squeezing in a parametric resonator by analyzing the noise spectral
density (NSD) in two channels in
quadrature of the slowly-varying variables based on Louisell's coupled mode method and by
making subsequent phase averages over the external ac drive, not via the
analysis of stochastic differential equations. 
Again, the results were limited to resonance when half the pump frequency is
equal to the natural frequency of a single-degree-of-freedom parametric
resonator.

There have been attempts at increasing thermal noise squeezing
beyond the $-6$~dB level using several different systems and methods such as
two-mode coupling in nanomechanical resonators
\cite{mahboob2014two,patil2015thermomechanical, pontin2016dynamical,singh2018motion}, multi-mode coupling
\cite{zhao2021two}, feedback \cite{vinante2013feedback, poot2015deep, sonar2018strong, mashaal2024strong}, and
nonlinear squeezing \cite{almog2007noise, huber2020spectral}.
All methods used are approximate solutions adapted to each problem, none
of them being general.
Instead, what we propose here is a theoretical method that could be applied to
all linear parametric amplifier systems without feedback.
In addition to that, our method can be applied to nonlinear squeezing if the
added noise can be considered as a small perturbation.
In this case, the response to noise will be given by a
parametrically driven system in which the pump is given by a stable limit cycle
solution of the unperturbed dynamical system (i.e. without added noise) \cite{wiesenfeld1985noisy}.
The only systems that cannot be covered by our present analysis are those with
feedback. 

We propose a theoretical model based on stochastic differential equations to
explain sub-threshold thermal noise squeezing such as first observed by Rugar
and Grütter in a classical parametrically driven resonator.
In the case of single degree-of-freedom
parametrically-driven harmonic oscillators, this behavior is linked to 
the constraint imposed on the Floquet multipliers (FMs) by Liouville's formula
\cite{teschl2024ordinary}.
It imposes that the FMs can either be a complex conjugate pair
with a magnitude less than or equal to 1 or they can be two different real
numbers whose product is less than or equal to 1. 
The squeezing is enhanced when the Floquet exponents are real, leading to two
different effective dissipation rates. 

We start our theoretical model of noise squeezing analyzing the
frequency-domain response  of the  parametrically modulated
resonator to added white noise. 
The response consists of  elastic scattering and parametric up and 
down conversions.
The elastic response is a consequence of the time translation invariant part
of the Green's function.
The other responses arise from the non symmetric part of the Green's function.
Subsequently, we split the resonator response into real and imaginary components.
Afterwards, we obtain the average fluctuations of the real and imaginary parts
and the correlation between them.
For a generic pump phase, we obtain these statistical averages from a bivariate
Gaussian distribution with correlation.
By an appropriate change of coordinates, one can transform from this
distribution to a new distribution that is a product of two univariate Gaussian
distributions.
From the latter distributions, we obtain the length and width of the
distribution representing data points obtained in a lockin amplifier (LIA)
\cite{rugar91}.
Initially, we develop the approximate theory of noise squeezing based on 
the first-order averaging method and, later, we also develop an exact theory of noise squeezing based on Floquet
theory.
With this approach we obtained very deep squeezing of fluctuations, near $-40~$dB, better than any classical result in the literature.
This latter approach can be applied to multiple coupled parametrically
modulated resonators with added noise.

The remainder of this paper is organized as follows:
in Sec.~\ref{model} we state our problem and develop the analytical model of
noise squeezing based on the first-order averaging approximation (in
Subsec.~\ref{thermal}),
and on Floquet theory (in Subsec.~\ref{thermalF}).
In Sec.~\ref{numerical} we present and discuss our results.
In Sec.~\ref{conclusion} we draw our conclusions.
\section{Model and Analysis}
\label{model}
In this section, we develop theoretical models to explain the phenomenon
of classical parametric noise squeezing using two different methods: 
one using the first-order averaging approximation and
the other, exact (in principle) and more general, using Floquet theory.

We initially investigate the effect of added white noise on the
single-degree-of-freedom parametrically-driven
resonator \cite{batista2011cooling}.
Its stochastic differential equation can be written as
\begin{equation}
    \ddot x=-x-\gamma\dot x+F_p\cos(2\omega t+\varphi_p)\;x+r(t),
    \label{parampNoise}
\end{equation}
where $\gamma$ is the dissipation rate, $F_p$ is the pump amplitude, $\omega$ is
half the pump frequency, $\varphi_p$ is the pump phase, and $r(t)$ is a Gaussian white noise that satisfies the statistical averages
$\langle r(t)\rangle=0$
and $\langle r(t)r(t')\rangle= 2 D\delta(t-t')$, where $D$ is the noise level.

The second model we study is a parametric resonator coupled to a harmonic
resonator. 
It is the linearized version of the model experimentally investigated in
Refs.~\cite{singh2018motion, singh2020giant}.
Its dynamics is given by the following equations of motion
\beq
\begin{aligned}
\ddot x &=-\gamma_1 \dot x-[1-F_p\sin(2\omega t)]x+\beta_1 y+r_1(t),\\
\ddot y &= -\gamma_2\dot y-\omega_2^2y+\beta_2 x+r_2(t),
\end{aligned}
\label{2ModeParamp}
\eeq
where $r_1(t)$ and $r_2(t)$ are 
Gaussian white noises that satisfy the statistical averages
$\langle r_1(t)\rangle=\langle r_2(t)\rangle=0$
and $\langle r_i(t)r_j(t')\rangle= 2 D\delta_{ij}\delta(t-t')$ for $i,j=1,2$,
with $\delta_{ij}=1$ if $i=j$ and $0$ otherwise.

The third and most general model we investigate consists of multiple coupled damped parametrically-driven oscillators with
added white noise.
Its Langevin equation is given by
\beq
{\bf M}\ddot X+{\bf\Gamma} \dot X+{\bf K}X=\left[{\bf C}e^{2i\omega t}+{\bf C^*}e^{-2i\omega t}\right]X+R(t),
\label{cpo_noise}
\eeq
where $X\in\mathbb R^N$, $\bf M$ and $\bf\Gamma$ are positive definite diagonal
$N\times N$ matrices, ${\bf K}\in\mathbb R^{N\times N}$ is the elastic coupling $N\times N$ matrix which is symmetric, and ${\bf C}\in\mathbb
C^{N\times N}$.
The element  ${\bf M}_{ii}=m_i$ is the mass (effective mass) of each
resonator (mode). 
In general, the matrices ${\bf M}$, ${\bf \Gamma}$, ${\bf K}$, and ${\bf C}$
cannot be simultaneously diagonalized.
If they can, one could decouple the oscillations into a supperpostion of normal
modes, turning the problem in a much simpler format, with equations similar to
Eq.~\eqref{2ModeParamp} for each normal mode.
In this case, one would necessarily be limited to $-6~$dB lower limit for
squeezing.
Here we are interested in the most general case in which the coefficient
matrices cannot be diagonalized simultaneously.
\subsection{Noise squeezing in the 1st-order averaging approximation}
\label{thermal}
The response in frequency space of the parametrically driven resonator to added
white noise \cite{batista2022gain} is given by
\beq
\begin{aligned}
\tilde x(\nu) &=\tilde G_0(\nu)\tilde r(\nu)-\frac{\beta}{2\omega}
\left[ 
\frac{e^{-i\varphi_p}\tilde r(\nu-2\omega)}{[\rho_-+i(\nu-\omega)][\rho_+ +i(\nu-\omega)]}
+\frac{e^{i\varphi_p}\tilde r(\nu+2\omega)}{[\rho_-+i(\nu+\omega)][\rho_++i(\nu+\omega)]}
\right]\\
&=\tilde G_0(\nu)\tilde r(\nu)+\Gamma(\nu)\tilde r(\nu-2\omega)
+\Gamma^*(-\nu)\tilde r(\nu+2\omega),
\label{noise_response}
\end{aligned}
\eeq
where the symbol $^*$ denotes complex conjugation.
Here, we use the shorthand notations
\begin{align}
\Gamma(\nu)
&=-\frac{\beta}{2\omega}\frac{e^{-i\varphi_p}}{[\rho_-+i(\nu-\omega)][\rho_++i(\nu-\omega)]},\\
\tilde G_0(\nu)&
\approx
\frac{1}{2\omega}\left[
\frac{\delta+\nu-\omega+i\gamma/2}{[\rho_-+i(\nu-\omega)][\rho_++i(\nu-\omega)]}
+\frac{\delta-\nu-\omega-i\gamma/2}{[\rho_-+i(\nu+\omega)][\rho_++i(\nu+\omega)]}
\right],
\label{G_0}
\end{align}
in which $\rho_\pm=-\gamma/2\pm\kappa$ are the Floquet exponents
in the first-order averaging approximation,
with $\kappa=\sqrt{\beta^2-\delta^2}$, $\beta=-F_p/4\omega$, 
$\delta=\Omega/2\omega$, and $\Omega=1-\omega^2$ \cite{batista2012heating}.
Near the onset of instability, at $\nu=\omega$, we have
\beq
\tilde G_0(\omega)
\approx\frac{1}{2\omega}\left[
\frac{\delta+i\gamma/2}{\rho_-\rho_+}
+\frac{\delta-2\omega-i\gamma/2}{(\rho_-+2i\omega)(\rho_++2i\omega)}
\right]
\approx\frac{\delta+i\gamma/2}{2\omega\rho_-\rho_+}.
\label{G_0om}
\eeq
Here, we use the following notation for Fourier transforms
\[
\tilde f(\nu)=\int_{-\infty}^\infty e^{i\nu t} f(t)\,dt.
\]

We are now going to investigate the thermal noise squeezing phenomenon that
occurs in the frequency domain of parametrically driven oscillators with added
white noise.
In order to do that we calculate the real and imaginary parts of $\tilde
x(\nu)$, which we call $\tilde x'$ and $\tilde x''$, respectively.
From Eq.~\eqref{noise_response}, we find
\beq
\begin{aligned}
    \tilde x'(\nu) &=\tilde G_0'(\nu)\tilde r'(\nu)-\tilde G_0''(\nu)\tilde r''(\nu)
   + 
    \Gamma'(\nu)\tilde r'(\nu-2\omega)-\Gamma''(\nu)\tilde r''(\nu-2\omega)
    \\&+\Gamma'(-\nu)\tilde r'(\nu+2\omega)+\Gamma''(-\nu)\tilde r''(\nu+2\omega),\\
    \tilde x''(\nu) &=\tilde G_0'(\nu)\tilde r''(\nu)+\tilde G_0''(\nu)\tilde r'(\nu)
    +
    \Gamma'(\nu)\tilde r''(\nu-2\omega)+\Gamma''(\nu)\tilde r'(\nu-2\omega)
\\&+\Gamma'(-\nu)\tilde r''(\nu+2\omega)-\Gamma''(-\nu)\tilde r'(\nu+2\omega),
\end{aligned}
\label{xp_xpp}
\eeq
where the real and imaginary parts of $\tilde r$ are $\tilde r'$ and $\tilde
r''$, respectively.
Using the parity properties of the Fourier transform of a real function
\beq
\begin{aligned}
    \tilde r'(\nu)&=\tilde r'(-\nu),\\
    \tilde r''(\nu) &=-\tilde r''(-\nu),
\end{aligned}
\eeq
and the following statistical averages of white noise in the frequency domain:
\beq
\begin{aligned}
\langle \tilde r'(\nu)\tilde r'(\nu')\rangle &=2\pi
D\left[\delta(\nu-\nu')+\delta(\nu+\nu')\right],\\
\langle \tilde r'(\nu)\tilde r''(\nu')\rangle &=0,\\
\langle \tilde r''(\nu)\tilde r''(\nu')\rangle &=2\pi
D\left[\delta(\nu-\nu')-\delta(\nu+\nu')\right],
\end{aligned}
\eeq
we obtain the dispersions in quadrature of the response of the parametric
resonator to added noise, when $\nu\neq\omega$.
They are given by
\beq
\begin{aligned}
\sigma_c^2(\nu)&=\lim_{\Delta\nu\rightarrow 0^+}\int^{\nu+\Delta\nu}_{\nu-\Delta\nu} 
\langle \tilde x'(\nu)\tilde x'(\nu')\rangle\; d\nu'=
2\pi D\left[|\tilde
G_0(\nu)|^2+|\Gamma(\nu)|^2+|\Gamma(-\nu)|^2\right],\\
\sigma_s^2(\nu)&=\lim_{\Delta\nu\rightarrow 0^+}\int^{\nu+\Delta\nu}_{\nu-\Delta\nu} 
\langle \tilde x''(\nu)\tilde x''(\nu')\rangle\; d\nu'=
2\pi D\left[|\tilde
G_0(\nu)|^2+|\Gamma(\nu)|^2+|\Gamma(-\nu)|^2\right].
\end{aligned}
\label{sigma2_cs_nu}
\eeq
Here, $\sigma_c$ is the dispersion of the cosine quadrature component and 
$\sigma_s$ is the dispersion of the sine quadrature component.
We also obtain that there is no correlation between $\tilde x'(\nu)$ and
$\tilde x''(\nu)$ when $\nu\neq\omega$, that is
\[
\lim_{\Delta\nu\rightarrow 0^+}\int^{\nu+\Delta\nu}_{\nu-\Delta\nu} 
\langle \tilde x'(\nu)\tilde x''(\nu')\rangle\; d\nu'= 0.
\]

Before we proceed with our theoretical development, we verify that the results
we obtain for the NSDs $S_{\tilde x'}$ and
$S_{\tilde x''}$ of $\tilde x'$ and $\tilde x''$, respectively, are consistent
with the NSD of $\tilde x$. 
That is, we verify below that $S_{\tilde x}=S_{\tilde x'}+S_{\tilde x''}$ is 
indeed correct as expected.
The NSD $S_{\tilde x}$ is defined \cite{batista2022gain} as
\beq
S_{\tilde x}(\nu)=\lim_{\Delta\nu\rightarrow 0^+}\int^{\nu+\Delta\nu}_{\nu-\Delta\nu} 
\dfrac{\langle \tilde x(-\nu)\tilde x(\nu')\rangle}{2\pi}\; d\nu'.
\label{S_N}
\eeq
With the help of Eq.~\eqref{noise_response} and Eq.~\eqref{sigma2_cs_nu}, when $\nu\neq\omega$, 
it can be written as
\beq
S_{\tilde x}(\nu)=\frac{\sigma_c^2+\sigma_s^2}{2\pi}=2D\left[|\tilde
G_0(\nu)|^2+|\Gamma(\nu)|^2+|\Gamma(-\nu)|^2\right].
\label{S_Nnu}
\eeq
We note that this result is in agreement with the NSD obtained
and experimentally confirmed in Ref. \cite{batista2022gain} in an analog
electronic circuit and in a mechanical resonator in Ref.~\cite{miller2020spectral}.
At $\nu=\omega$, we obtain the two dispersions in quadrature and the
correlation to be given by
\beq
\begin{aligned}
\sigma_c^2(\omega)&=\lim_{\Delta\nu\rightarrow 0^+}\int^{\omega+\Delta\nu}_{\omega-\Delta\nu} 
\langle \tilde x'(\omega)\tilde x'(\nu')\rangle\; d\nu'\\
&=
2\pi D\left\{|\tilde G_0(\omega)|^2+|\Gamma(\omega)|^2+|\Gamma(-\omega)|^2
+2\Re\{\tilde G_0(\omega)\Gamma(\omega)\}\right\},
\end{aligned}
\eeq
\beq
\begin{aligned}
\sigma_s^2(\omega)&=\lim_{\Delta\nu\rightarrow 0^+}\int^{\omega+\Delta\nu}_{\omega-\Delta\nu} 
\langle \tilde x''(\omega)\tilde x''(\nu')\rangle\; d\nu'\\
&=2\pi D\left\{|\tilde G_0(\omega)|^2+|\Gamma(\omega)|^2+|\Gamma(-\omega)|^2
-2\Re\{\tilde G_0(\omega)\Gamma(\omega)\}
\right\},
\end{aligned}
\eeq
and
\beq
\begin{aligned}
\sigma_{cs}(\omega)&=\lim_{\Delta\nu\rightarrow 0^+}\int^{\omega+\Delta\nu}_{\omega-\Delta\nu} 
\langle \tilde x'(\omega)\tilde x''(\nu')\rangle\; d\nu'
=4\pi D \Im\left\{\tilde G_0(\omega)\Gamma(\omega)\right\}.
\label{sigma_cs_om}
\end{aligned}
\eeq
Experimental data points sampled (with a long enough sample time interval) by a
LIA are statistically independent random variates generated by a Gaussian
probability distribution with zero mean and these corresponding dispersions and
correlation.
As the pump amplitude is increased near the instability threshold, the
correlation grows giving rise to what is known in 
the literature as thermal noise squeezing \cite{rugar91}.
As $|\Gamma(-\omega)|<<|\Gamma(\omega)|$, we can further simplify
these expressions.
More simply, we can write
\beq
\begin{aligned}
\sigma_c^2(\omega)
&\approx2\pi D\left\{
|\tilde G_0(\omega)|^2+|\Gamma(\omega)|^2
-\frac{\beta}{\omega\rho_-\rho_+} 
\left[\tilde G_0'(\omega)\cos\varphi_p-\tilde G_0''(\omega)\sin\varphi_p\right]
\right\}\\
&= \frac{\pi
D}{2(\omega\rho_-\rho_+)^2}\left(\delta^2+\gamma^2/4+\beta^2-2\beta\delta\cos\varphi_p+\beta\gamma\sin\varphi_p\right)
\\
&= \frac{\pi
D}{2\omega^2}\frac{\delta^2+\gamma^2/4+\beta^2-2\beta\delta\cos\varphi_p+\beta\gamma\sin\varphi_p}{\left(\gamma^2/4-\kappa^2\right)^2},
\end{aligned}
\label{sigma2_c_om_approx}
\eeq

\beq
\begin{aligned}
\sigma_s^2(\omega)
&\approx2\pi D\left\{
|\tilde G_0(\omega)|^2+|\Gamma(\omega)|^2
+\frac{\beta}{\omega\rho_-\rho_+} 
\left[\tilde G_0'(\omega)\cos\varphi_p-\tilde G_0''(\omega)\sin\varphi_p\right]
\right\}\\
&= \frac{\pi
D}{2(\omega\rho_-\rho_+)^2}\left(\delta^2+\gamma^2/4+\beta^2+2\beta\delta\cos\varphi_p-\beta\gamma\sin\varphi_p\right)
\\
&= \frac{\pi D}{2\omega^2}
\frac{\delta^2+\gamma^2/4+\beta^2+2\beta\delta\cos\varphi_p-\beta\gamma\sin\varphi_p}{(\gamma^2/4-\kappa^2)^2},\\
\end{aligned}
\label{sigma2_s_om_approx}
\eeq
\beq
\hspace{-5cm}\sigma_{cs}(\omega)
\approx
-\frac{\pi
D\beta}{\omega^2(\gamma^2/4-\kappa^2)^2}\left(\frac\gamma2\cos\varphi_p-\delta\sin\varphi_p\right).
\label{sigma2_cs_om_approx}
\eeq
Notice that below the instability threshold ($\beta^2\leq \gamma^2/4+\delta^2$),
one can show that both dispersions are positive.
Assuming that $0<|\beta|<<1$, we then obtain
\beq
\sigma_c\sigma_s\approx2\pi D|\tilde G_0(\omega)|^2.
\eeq
For the important special case of $\delta=0$ and $\varphi_p=-\pi/2$, we find
\beq
\begin{aligned}
\sigma_c^2(\omega)&= \frac{\pi D}{2\omega^2}\frac1{\left(\frac\gamma2+\beta\right)^2},\\
\sigma_s^2(\omega)&= \frac{\pi D}{2\omega^2}\frac1{\left(\frac\gamma2-\beta\right)^2},\\
\sigma_{cs}(\omega) &=0.
\end{aligned}
\eeq
Further, notice that there is noise squeezing in the dispersions only at
$\nu=\omega$.
With the help of Eq.~\eqref{sigma_cs_om}, the noise spectral density at $\omega$
is given by
\beq
S_{\tilde x}(\omega)
=2D\left[|\tilde G_0(\omega)|^2+
|\Gamma(\omega)|^2+|\Gamma(-\omega)|^2\right]
\approx\frac{D}{2\omega^2}\frac{\delta^2+\gamma^2/4+\beta^2}{\left(\gamma^2/4-\kappa^2\right)^2}.
\eeq
Hence, we see that $S_{\tilde x}(\nu)$ is a continuous function of $\nu$, unlike
$\sigma_c(\nu)$ and $\sigma_s(\nu)$ which present a discontinuous behavior at
$\nu=\omega$.
We believe that the discontinuities at $\nu=\omega$ of $\sigma_c(\nu)$ and
$\sigma_s(\nu)$ are due to the fact that we assumed that the noise is completely
uncorrelated in time and, consequently, in frequency as well.
In reality, when $\omega$ is very close to $\omega'$, one should get
$\langle\tilde r(\omega)\tilde r(\omega')\rangle\neq 0$.
How close these frequencies have to be so that the correlation is appreciable
depends on the physical process that generates the noise $r(t)$.

\subsection{Noise squeezing using Floquet theory}
\label{thermalF}
According to Ref.~\cite{batista2024amplification}, the response in frequency
space of the parametrically driven resonator to added white noise is given by
\beq
\tilde X(\nu)\approx{\bf\tilde G_0}(\nu)\tilde R(\nu)+{\bf G_+}(\nu)\tilde R(\nu-2\omega)
+{\bf G_+^*}(-\nu)\tilde R(\nu+2\omega),
\label{fundEqMat}
\eeq
where $\tilde X$ and $\tilde R\in \mathds{C}^{2N}$ and the matrices $\mathbf{\tilde
G_0}$ and $\bf G_+$ $\in \mathds{C}^{2N\times 2N}$.
Here, these matrices (the generalized Green's function) are given by
\beq
\begin{aligned}
    \mathbf{\tilde G_0}(\nu)&\approx
-p_1^*\left[B+i(\nu-\omega)I\right]^{-1}q_1-p_1\left[B+i(\nu+\omega)I\right]^{-1}q_1^*\\
&=-\sum_k\frac{p_1^*|v_{k}\rangle\langle v_{k}|q_1}{\rho_k+i(\nu-\omega)}
-\sum_k\frac{p_1|v_{k}\rangle\langle v_{k}|q_1^*}{\rho_k+i(\nu+\omega)},\\
\bf G_+(\nu) &= -p_1^*\left[B+i(\nu-\omega)I\right]^{-1}q_1^*
= -\sum_k\frac{p_1^*|v_{k}\rangle\langle v_{k}|q_1^*}{\rho_k+i(\nu-\omega)}.\\
\end{aligned}
\label{ABcoefs}
\eeq
The vectors $|v_k\rangle$ are Floquet eigenvectors with corresponding
Floquet exponents $\rho_k$.
Note that according to Floquet's theorem, 
the fundamental matrix of coherent evolution of Eq.\eqref{parampNoise},
i.e. without the added noise, can be written as $\Phi(t)=P(t)e^{Bt}$, where
$P(t+T)=P(t)$ is a periodic matrix with $T=2\pi/\omega$. 
Therefore, we can write the Fourier series expansions of the periodic matrices
\beq
\begin{aligned}
P(t)=\sum_{n=-\infty}^\infty p_{n}e^{2in\pi t/T},\\
Q(t)=\sum_{n=-\infty}^\infty q_{n}e^{2in\pi t/T},
\end{aligned}
\label{PQ}
\eeq
where $Q(t)=P^{-1}(t)$.

\subsubsection{Two-dimensional case}
For the case of a single degree-of-freedom parametric resonator with added
noise
we have
\beq
\tilde x(\nu)\approx\tilde \mg_0(\nu)\tilde r(\nu)+\mg_+(\nu)\tilde r(\nu-2\omega)
+\mg_+^*(-\nu)\tilde r(\nu+2\omega),
\label{fundEq2}
\eeq
where the coefficients 
$\tilde \mg_0(\nu)=\langle1|\mathbf{\tilde G_0}(\nu)|2\rangle$
and $\mg_+(\nu)=\langle1|\bf G_+(\nu)|2\rangle$.
Note that in this subsection, we replaced the notation $\tilde X_1$ by $\tilde x$ and $\tilde R_1$ by $\tilde r$.

Similarly to Eq.~\eqref{xp_xpp}, we find
\beq
\begin{aligned}
    \tilde x'(\nu) &=\tilde\mg_0'(\nu)\tilde r'(\nu)-\tilde\mg_0''(\nu)\tilde r''(\nu)+\left[
    \mg_+'(\nu)\tilde r'(\nu-2\omega)-\mg_+''(\nu)\tilde r''(\nu-2\omega)
    \right.\\
    +&\left.
    \mg_+'(-\nu)\tilde r'(\nu+2\omega)+\mg_+''(-\nu)\tilde r''(\nu+2\omega)
    \right],\\
    \tilde x''(\nu) &=\tilde \mg_0'(\nu)\tilde r''(\nu)+\tilde \mg_0''(\nu)\tilde r'(\nu)+\left[
    \mg_+'(\nu)\tilde r''(\nu-2\omega)+\mg_+''(\nu)\tilde r'(\nu-2\omega)
    \right.\\
    +&\left.
    \mg_+'(-\nu)\tilde r''(\nu+2\omega)-\mg_+''(-\nu)\tilde r'(\nu+2\omega)
    \right].
\end{aligned}
\label{xp_xpp2}
\eeq
Hence, we obtain the dispersions in quadrature of the response of the parametric
resonator to added noise, when $\nu\neq\omega$.
They are given by
\beq
\begin{aligned}
\sigma_c^2(\nu)&=\lim_{\Delta\nu\rightarrow 0^+}\int^{\nu+\Delta\nu}_{\nu-\Delta\nu} 
\langle \tilde x'(\nu)\tilde x'(\nu')\rangle\; d\nu'=
2\pi D\left\{|\tilde
\mg_0(\nu)|^2+|\mg_+(\nu)|^2+|\mg_+(-\nu)|^2\right\},\\
\sigma_s^2(\nu)&=\lim_{\Delta\nu\rightarrow 0^+}\int^{\nu+\Delta\nu}_{\nu-\Delta\nu} 
\langle \tilde x''(\nu)\tilde x''(\nu')\rangle\; d\nu'=
2\pi D\left\{|\tilde
\mg_0(\nu)|^2+|\mg_+(\nu)|^2+|\mg_+(-\nu)|^2\right\}.
\end{aligned}
\label{sigma2_cs_nuF}
\eeq
We also obtain that there is no correlation between $\tilde x'(\nu)$ and
$\tilde x''(\nu)$ when $\nu\neq\omega$, that is
\[
\lim_{\Delta\nu\rightarrow 0^+}\int^{\nu+\Delta\nu}_{\nu-\Delta\nu} 
\langle \tilde x'(\nu)\tilde x''(\nu')\rangle\; d\nu'= 0.
\]
With the help of the definitions of NSD given in Eq.~\eqref{S_N} and of
Eq.~\eqref{fundEq2}, when $\nu\neq\omega$, $S_{\tilde x}$ can be written as
\beq
S_{\tilde x}(\nu) = 2D\left\{|\tilde \mg_0(\nu)|^2+|\mg_+(\nu)|^2+|\mg_+(-\nu)|^2\right\}.
\eeq

When $\nu=\omega$, we obtain the two dispersions in quadrature and the
correlation to be given by
\begin{align}
\sigma_c^2(\omega)&=\lim_{\Delta\nu\rightarrow 0^+}\int^{\omega+\Delta\nu}_{\omega-\Delta\nu} 
\langle \tilde x'(\omega)\tilde x'(\nu')\rangle\; d\nu'\nn\\
&=2\pi D\left[|\tilde\mg_0(\omega)|^2+|\mg_+(\omega)|^2+|\mg_+(-\omega)|^2
+2\Re\left\{\tilde\mg_0(\omega)\mg_+(\omega)\right\}\right],
\label{sigma2_c_omF}\\
\sigma_s^2(\omega)&=\lim_{\Delta\nu\rightarrow 0^+}\int^{\omega+\Delta\nu}_{\omega-\Delta\nu} 
\langle \tilde x''(\omega)\tilde x''(\nu')\rangle\; d\nu'
\nn\\
&=2\pi D\left[|\tilde\mg_0(\omega)|^2+|\mg_+(\omega)|^2+|\mg_+(-\omega)|^2
-2\Re\left\{\tilde\mg_0(\omega)\mg_+(\omega)\right\}\right],
\label{sigma2_s_omF}\\
\sigma_{cs}(\omega)&=\lim_{\Delta\nu\rightarrow 0^+}\int^{\omega+\Delta\nu}_{\omega-\Delta\nu} 
\langle \tilde x'(\omega)\tilde x''(\nu')\rangle\; d\nu'=
4\pi D \Im\left\{\tilde\mg_0(\omega)\mg_+(\omega)\right\}.
\label{sigma_cs_omF}
\end{align}
\subsubsection{The $2N$-dimensional case}
According to Eq.~\eqref{fundEqMat}, the response in frequency
space of an $2N$-dimensional parametrically driven linear dynamical system to 
added white noise is given by
\beq
\tilde X_i(\nu)\approx\sum_j\left[{\bf\tilde G}_{0,ij}(\nu)\tilde R_j(\nu)
+{\bf G}_{+,ij}(\nu)\tilde R_j(\nu-2\omega)
+{\bf G}^*_{+,ij}(-\nu)\tilde R_j(\nu+2\omega)\right].
\label{fundEq3}
\eeq
We point out that from these $2N$ coordinates, $N$ are position and $N$ are
velocity coordinates.
In the notation we use, an odd index points to a position coordinate and an
even index points to a velocity coordinate.
For this case, we find
\beq
\begin{aligned}
    \tilde X'_i(\nu) &=\sum_j\bigg\{{\bf\tilde G}'_{0,ij}(\nu)\tilde
    R_j'(\nu)-{\bf\tilde G}''_{0,ij}(\nu)\tilde R_j''(\nu)+\left[
    {\bf G}'_{+,ij}(\nu)\tilde R_j'(\nu-2\omega)-{\bf G}''_{+,ij}(\nu)\tilde R_j''(\nu-2\omega)
    \right.\\
    +&\left.
    {\bf G}'_{+,ij}(-\nu)\tilde R_j'(\nu+2\omega)+{\bf G}''_{+,ij}(-\nu)\tilde R_j''(\nu+2\omega)
    \right]\bigg\},\\
    \tilde X''_i(\nu) &=\sum_j\bigg\{{\bf\tilde G}'_{0,ij}(\nu)\tilde
    R_j''(\nu)+{\bf\tilde G}_{0,ij}''(\nu)\tilde R_j'(\nu)
    +\left[
    {\bf G}'_{+,ij}(\nu)\tilde R_j''(\nu-2\omega)+{\bf G}''_{+,ij}(\nu)\tilde R_j'(\nu-2\omega)
    \right.\\
    +&\left.
    {\bf G}'_{+,ij}(-\nu)\tilde R_j''(\nu+2\omega)-{\bf G}''_{+,ij}(-\nu)\tilde R_j'(\nu+2\omega)
    \right]\bigg\}.
\end{aligned}
\label{xp_xpp3}
\eeq
When $\nu=\omega$, we obtain the dispersions in quadrature and the
correlations to be given by
\begin{align}
    \sigma_{i,c}^2(\omega)&=\lim_{\Delta\nu\rightarrow 0^+}\int^{\omega+\Delta\nu}_{\omega-\Delta\nu} 
\langle \tilde X_i'(\omega)\tilde X_i'(\nu')\rangle\; d\nu'\nn\\
&=2\pi D\sum_j\left[|{\bf\tilde G}_{0,ij}(\omega)|^2+|{\bf G}_{+,ij}(\omega)|^2
+|{\bf G}_{+,ij}(-\omega)|^2
+2\Re\left\{{\bf\tilde G}_{0,ij}(\omega){\bf G}_{+,ij}(\omega)\right\}\right],\\
\sigma_{i,s}^2(\omega)&=\lim_{\Delta\nu\rightarrow 0^+}\int^{\omega+\Delta\nu}_{\omega-\Delta\nu} 
\langle \tilde X_i''(\omega)\tilde X_i''(\nu')\rangle\; d\nu'\nn\\
&=2\pi D\sum_j\left[|{\bf\tilde G}_{0,ij}(\omega)|^2+|{\bf G}_{+,ij}(\omega)|^2
+|{\bf G}_{+,ij}(-\omega)|^2
-2\Re\left\{{\bf\tilde G}_{0,ij}(\omega){\bf G}_{+,ij}(\omega)\right\}\right],\\
\sigma_{i,cs}(\omega)&=\lim_{\Delta\nu\rightarrow 0^+}\int^{\omega+\Delta\nu}_{\omega-\Delta\nu} 
\langle \tilde X_i'(\omega)\tilde X_i''(\nu')\rangle\; d\nu'=
4\pi D \sum_j\Im\left\{{\bf\tilde G}_{0,ij}(\omega){\bf G}_{+,ij}(\omega)\right\},
\label{sigma_cs_omFG}
\end{align}
where we used
\beq
\begin{aligned}
\langle \tilde R_i'(\nu)\tilde R_j'(\nu')\rangle &=2\pi
D\delta_{ij}\left[\delta(\nu-\nu')+\delta(\nu+\nu')\right],\\
\langle \tilde R_i'(\nu)\tilde R_j''(\nu')\rangle &=0,\\
\langle \tilde R_i''(\nu)\tilde R_j''(\nu')\rangle &=2\pi
D\delta_{ij}\left[\delta(\nu-\nu')-\delta(\nu+\nu')\right].
\end{aligned}
\eeq
The most general case of dispersions where we take into account all the cross
correlations is given by
\beq
\begin{aligned}
    \sigma_{ij,c}^2(\omega)&=\lim_{\Delta\nu\rightarrow 0^+}\int^{\omega+\Delta\nu}_{\omega-\Delta\nu} 
\langle \tilde X_i'(\omega)\tilde X_j'(\nu')\rangle\; d\nu'\\
&=2\pi
D\sum_k\left[{\bf G}'_{0,ik}(\omega){\bf G}'_{0,jk}(\omega)
+{\bf G}''_{0,ik}(\omega){\bf G}''_{0,jk}(\omega)
+{\bf G}'_{+,ik}(\omega){\bf G}'_{+,jk}(\omega)
+{\bf G}''_{+,ik}(\omega){\bf G}''_{+,jk}(\omega)\right.\\
&\left.+\Re\left\{{\bm G}_{0,ik}(\omega){\bm G}_{+,jk}(\omega)+{\bm G}_{+,ik}(\omega){\bm G}_{0,jk}(\omega)\right\}\right],\\
\sigma_{ij,s}^2(\omega)&=\lim_{\Delta\nu\rightarrow 0^+}\int^{\omega+\Delta\nu}_{\omega-\Delta\nu} 
\langle \tilde X_i''(\omega)\tilde X_j''(\nu')\rangle\; d\nu'\\
&=2\pi
D\sum_k\left[{\bm G}'_{0,ik}(\omega){\bm G}'_{0,jk}(\omega)
+{\bm G}''_{0,ik}(\omega){\bm G}''_{0,jk}(\omega)
+{\bm G}'_{+,ik}(\omega){\bm G}'_{+,jk}(\omega)
+{\bm G}''_{+,ik}(\omega){\bm G}''_{+,jk}(\omega)\right.\\
&\left.-\Re\left\{{\bm G}_{0,ik}(\omega){\bm G}_{+,jk}(\omega)+{\bm G}_{+,ik}(\omega){\bm G}_{0,jk}(\omega)\right\}\right],\\
\sigma_{ij,cs}(\omega)&=\lim_{\Delta\nu\rightarrow 0^+}\int^{\omega+\Delta\nu}_{\omega-\Delta\nu} 
\langle \tilde X_i'(\omega)\tilde X_j''(\nu')\rangle\; d\nu'=
2\pi D \sum_k\left\{{\bm G}'_{0,ik}(\omega){\bm G}''_{0,jk}(\omega)
-{\bm G}''_{0,ik}(\omega){\bm G}'_{0,jk}(\omega)\right.\\
&\left.+{\bm G}'_{+,ik}(\omega){\bm G}''_{+,jk}(\omega)
-{\bm G}''_{+,ik}(\omega){\bm G}'_{+,jk}(\omega)+\Im\left[{\bm G}_{0,ik}(\omega){\bm G}_{+,jk}(\omega)
+{\bm G}_{+,ik}(\omega){\bm G}_{0,jk}(\omega)\right]\right\}.
\end{aligned}
\label{sigma_cs_omFG}
\eeq
In the Appendix B, we diagonalize the $2N\times2N$ covariance matrix 
whose entries are given above.
Based on Eq.~\eqref{eq:P_N}, the maximum amount of squeezing will be given by
the smallest eigenvalue of the covariance matrix.
The amount of squeezing thus obtained will be even higher than the previous 
method, but it will be harder to implement experimentally.

\section{Numerical Results and Discussion}
\label{numerical}
\begin{figure}[!ht]
    \centerline{\includegraphics[{scale=0.95}]{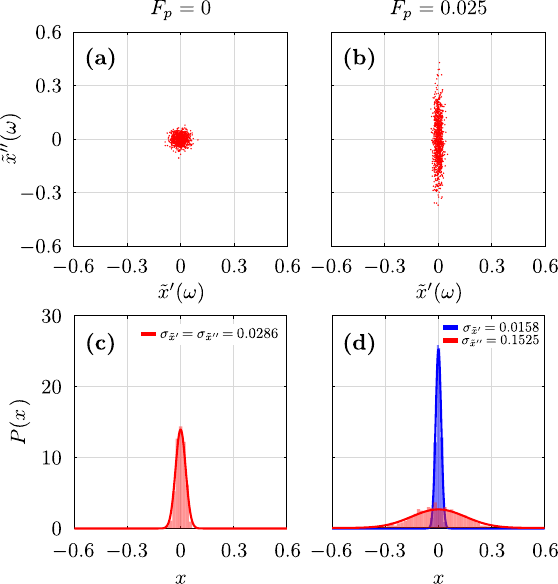}}
\caption{
In panels (a) and (b) we plot the clouds of data points of the Fourier
transforms $\tilde x(\omega)$. In panel (a) we have the equilibrium cloud with
no squeezing ($F_p=0$). In panel (b) we have squeezing with the pump amplitude
$F_p=0.025$ and pump phase $\varphi_p=-\pi/2$.
In all panels, we have $Q=65$ and $\omega=1$.
In panels (c) and (d), we plot the Gaussian distributions alongside the
histograms for the clouds from (a) and (b) panels, respectively.
In panel (c) we plot the equilibrium Gaussian distribution and in panel (d), we
plot the squeezed Gaussian distributions with zero mean
and dispersions given by Eq.~\eqref{sigma2_c_om_approx} for the $\tilde
x'(\omega)$ data points and by Eq.~\eqref{sigma2_s_om_approx} for the $\tilde
x''(\omega)$ data points. 
}
\label{fig:squeezing}
\end{figure}

\FloatBarrier
We plot in panel (a) of Fig.~\ref{fig:squeezing} the numerical data points in
the  $\tilde x'(\omega)$-$\tilde x''(\omega)$ plane with $F_p=0$ and $\omega=1$,
while in panel (b) we use $F_p=0.025$ and $\omega=1$. 
In panels (c) and (d), we plot the histograms of the data points
alongside the Gaussian distributions obtained from theory, corresponding
to the results of panels (a) and (b), respectively.
Each data point is a Fourier transform performed over a time series obtained
from the integration of Eq.~\eqref{parampNoise} with $\varphi_p=-\pi/2$. 
The integration time step is $dt=T/128$, where $T=2\pi/\omega$.
The first $512T$ of each time series are discarded so as transients die out and
the Fourier transform is performed over a time span of further $512T$.
There are $N=1000$ data points in this plot.
Not only we see squeezing in the elongated shape of the cloud of data points, 
but we also obtain quantitative agreement in fitting the normalized histograms
of the $\tilde x'(\omega)$ and $\tilde x''(\omega)$ data points in panel (d)
with the zero-mean Gaussian distributions with dispersions given by Eqs.
\eqref{sigma2_c_om_approx} and \eqref{sigma2_s_om_approx}.
We include the equilibrium distribution in panel (c) as a comparison.
In all data presented here we used the noise level $D=3.08\times10^{-8}$ 
(in dimensionless units) and the quality factor $Q=65$.
These values were obtained from the electronic circuit implementation of a
parametric resonator given in Ref.~\cite{batista2022gain}.
Here, we show that the stochastic differential equations model we developed 
explains well the squeezing data. 
Rugar and Grutter \cite{rugar91} plotted similar results obtained
experimentally, but their Gaussian distributions were only best fits from their
data.

\begin{figure}[!ht]
    \centerline{\includegraphics[{scale=0.4}]{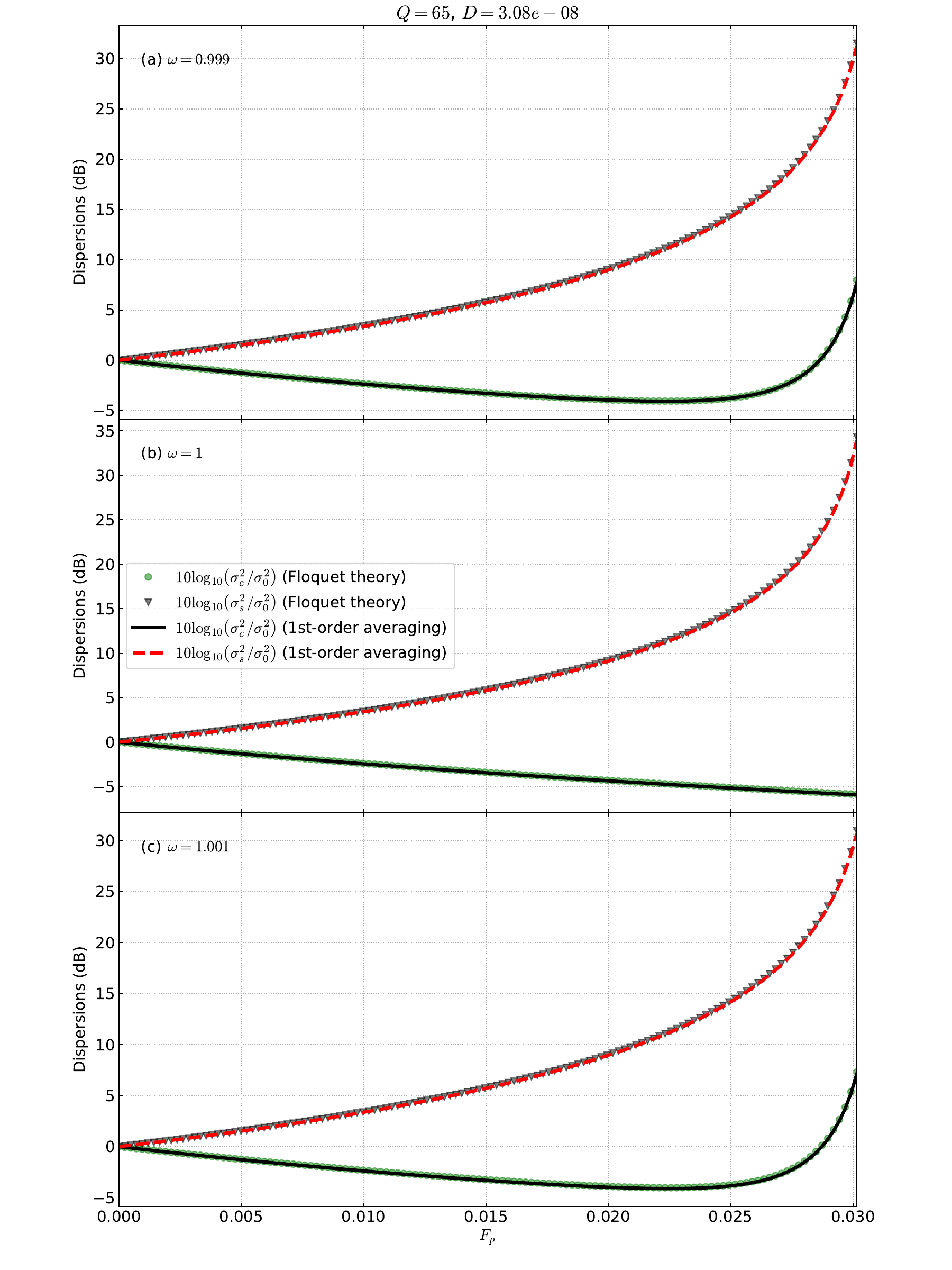}}
\caption{Comparison of the analytical (first-order averaging)
standard deviations given in Eqs.~\eqref{sigma2_c_om_approx}-\eqref{sigma2_s_om_approx}
of the real and imaginary parts of $\tilde x(\omega)$ with the corresponding
Floquet theory results given in Eqs.~\eqref{sigma2_c_omF}-\eqref{sigma2_s_omF}. 
The results are plotted in decibels relative to the equilibrium value with zero
pump.
In all cases $\varphi_p=-\pi/2$.
(a) At $\omega=0.999$, as the pump amplitude $F_p$ is increased up to the
instability threshold, one of the quadratures is squeezed down to a minimum
above $-6~$dB and then it starts growing again.
(b) At $\omega=1$ one sees that when the pump amplitude $F_p$ is increased up to
the instability threshold one of the quadratures is squeezed down to $-6$~dB in
amplitude while the other quadrature increases without bounds.
(c) At $\omega=1.001$ one sees basically the same behavior as in panel (a).}
\label{fig:stds}
\end{figure}

We plot in Fig.~\ref{fig:stds} the results for the dispersions in decibels
relative to the dispersion at zero pump as a function of pump amplitude.
One sees that the agreement between analytical (1st-order averaging
approximation) as presented in Eqs.~\eqref{sigma2_c_om_approx}-\eqref{sigma2_s_om_approx}
{} and Floquet theory results from
Eqs.~\eqref{sigma2_c_omF}-\eqref{sigma_cs_omF} is excellent. 
No correlation is taken into account here.
In panel (a) $\omega=0.999$. As the pump amplitude $F_p$ is increased up to the
instability threshold one of the quadratures is squeezed down to a minimum above
$-6~$dB and then it starts growing again.
This type of behavior was observed experimentally in \cite{mahboob2014two} and
in \cite{singh2018motion}.
In panel (b) $\omega=1$. There is a reduction of fluctuations in one quadrature and
an increase of fluctuations in the other quadrature as $F_p$ increases and
one nears the instability threshold.
The lower limit of deamplification is reached at threshold is $-6~$dB. 
There is basically no distinction between the results.
In panel (c) $\omega=1.001$.
One sees essentially the same behavior as in panel (a).
This might show that the squeezing phenomenon occurs in a very narrow frequency range
around resonance, but here we are not considering the effect of correlation 
as derived in Eqs.~\eqref{sigma2_cs_om_approx} and 
\eqref{sigma_cs_omF}, where 
it arises with detuning even when the pump is a sine function.
To obtain the Floquet exponents necessary to calculate the dispersions in this
figure, we used Scipy's Radau ODE integration method in the solve\_ivp solver.
We achieved more accurate results with this implicit method that is capable of
handling stiff differential equations than explicit  methods such as the
Runge-Kutta integration method of 5(4) order, which are generally unsuitable
for obtaining solutions of stiff differential equations.

\begin{figure}[!ht]
\centerline{\includegraphics[{scale=0.4}]{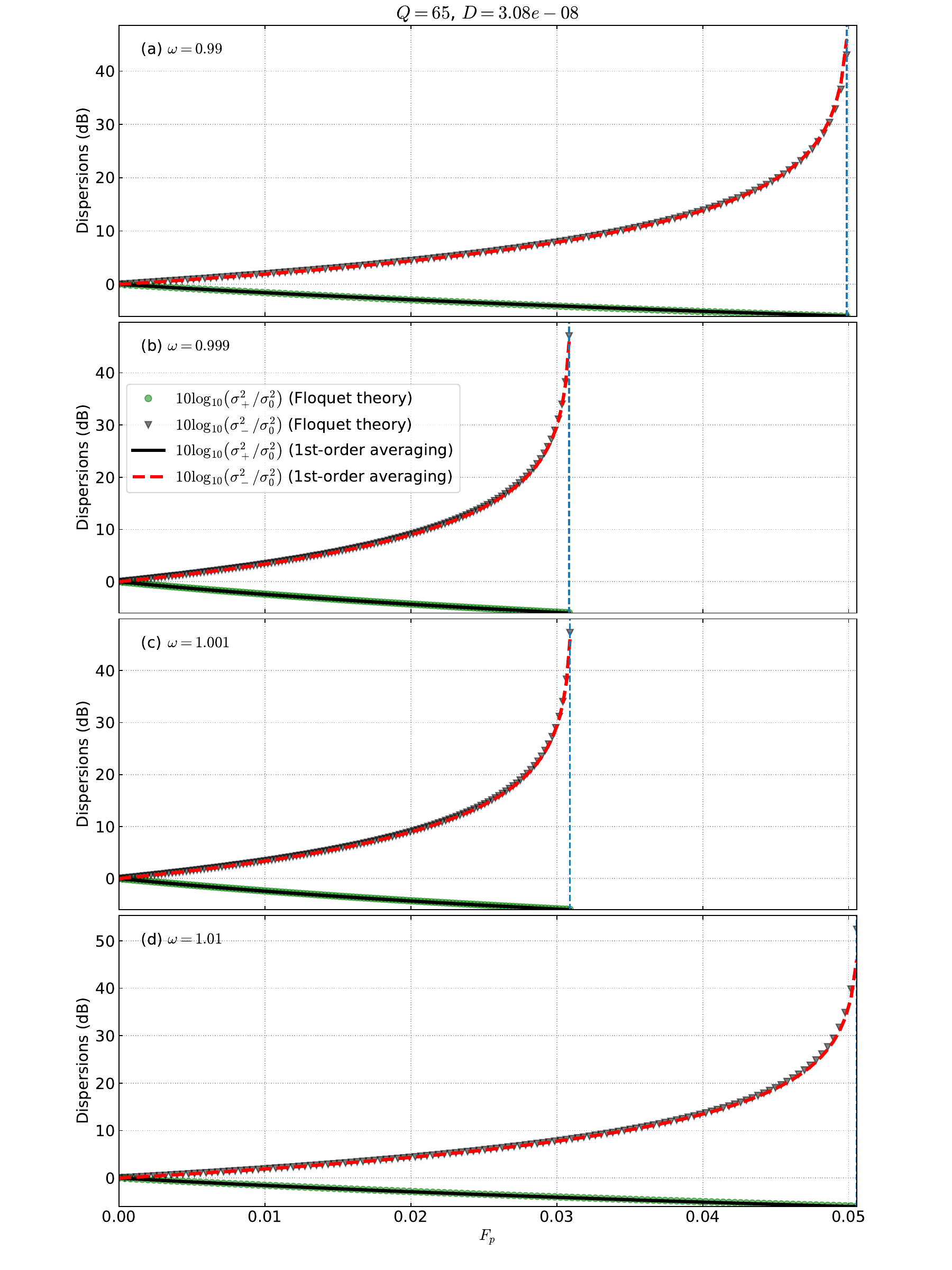}}
\caption{Comparison of the analytical (first-order averaging) standard of the
diagonalized dispersions (in the panel with no correlation) with the
corresponding Floquet theory result.
The diagonalization process is described in the Appendix A.
In panels (a)-(d), the $-6$~dB squeezing limit (at the bottom axes) can be
reached near the instability threshold (depicted by the vertical dashed lines)
even with detuning. 
The results for squeezing in panels (b) and (c) are in marked contrast with what
happens for the same parameters in Figs. \ref{fig:stds}(a) and
\ref{fig:stds}(c), respectively, where correlation was not taken into account.
In panels (a) and (d) the $-6$~dB limit can be reached even with a detuning 10
times larger than in panels (b) and (c).}
\label{fig:si_uv}
\end{figure}

In Fig. \ref{fig:si_uv}, we revisit the problem of squeezing in the single
parametric resonator, but now taking into consideration the effect of correlation. 
When there is correlation, the cloud of data points is no longer aligned with
the $\tilde x'(\omega)$ and $\tilde x''(\omega)$ as shown in Fig. \ref{fig:squeezing}.
It becomes tilted and the bivariate Gaussian distribution is no longer a product
of two uncorrelated Gaussian distributions of these two statistical variates.
In the Appendix A we show how to transform from one bivariate Gaussian
probability distribution with correlation to the diagonalized uncorrelated 
product of two univariate Gaussian distributions.
With this diagonalized form of distribution, we obtained the width and length of the squeezed cloud.
Along one principal axis, we have a dispersion $\sigma_+^2$ and along
the other we have $\sigma_-^2$ as given in Eq.~\eqref{eq:dispersions} with the
help of Eq.~\eqref{la_pm}.
The sigmas are given by
Eqs.~\eqref{sigma2_c_om_approx}-\eqref{sigma2_cs_om_approx} in first-order
averaging and by Eqs.~\eqref{sigma2_c_omF}-\eqref{sigma_cs_omF} in Floquet
theory.
These diagonalized dispersions are the correct measures of squeezing. 
One can see that the  results shown in panels (b) $\omega=0.999$ and (c)
$\omega=1.001$ are starkly different from the corresponding results shown in
panels \ref{fig:stds}(a) and \ref{fig:stds}(c).
We see here that the squeezing did not disappear as one might conclude from
hastily analyzing the results of Fig. \ref{fig:stds}.
In panels (a) and (d), we show that even with 10 times more detuning one can
reach the $-6$~dB lower limit of squeezing.
Hence, these results show that squeezing can be far more broad band than previously
thought in the literature \cite{rugar91, cleland2005thermomechanical,
mahboob2014two}.

\begin{figure}[!ht]
\centerline{\includegraphics[{scale=0.8}]{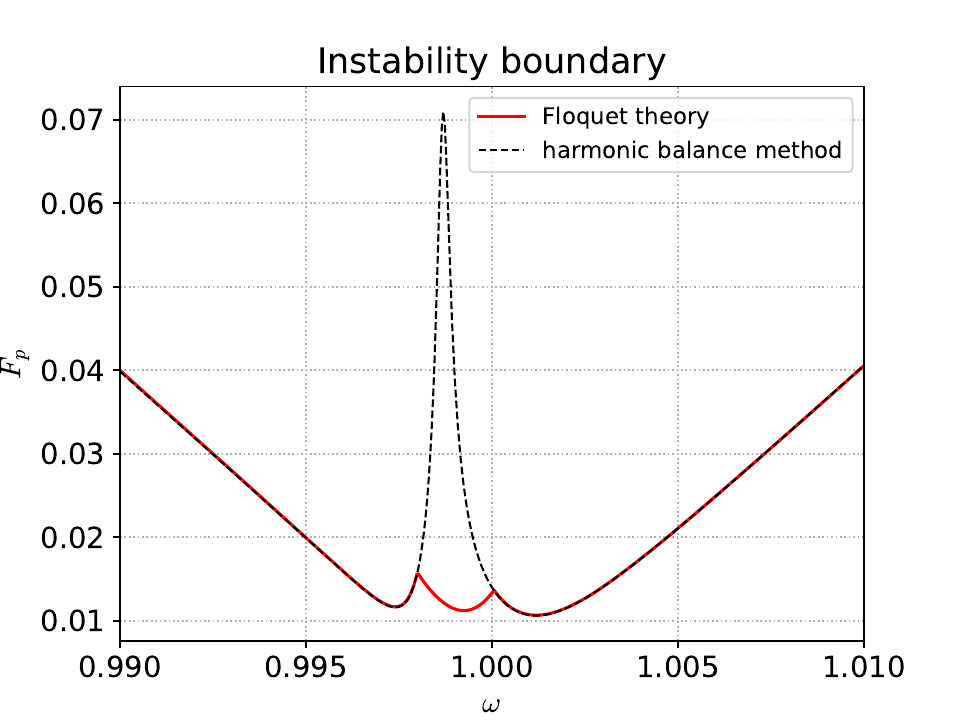}}
\caption{Instability boundary of the parametric-harmonic coupled resonators as
described in Eqs. \eqref{2ModeParamp} when $r_1(t)=r_2(t)=0$.
Below the threshold line, the quiescent solution is stable, whereas above it
the quiescent solution is unstable.
In the central region of the figure, the threshold line predicted by the
harmonic balance or averaging method is very different from the threshold
predicted by Floquet theory.
In this region, a Hopf bifurcation occurs at the red line. 
There, two complex Floquet multipliers (FMs) have unit magnitude, while the two
other multipliers have magnitudes less than one.
The remainder regions of the threshold line, depicted by the red black dashed line, correspond
to a period-doubling bifurcation where only one multiplier is equal to -1.
}
\label{fig:boundary}
\end{figure}

In Fig. \ref{fig:boundary} we show the transition line between the region of
$(\omega, F_p)$ parameter space where the quiescent solution is stable and the
region where it is unstable in the coupled parametric-harmonic resonators
model. 
Along most of this transition line there is agreement between the results of
harmonic balance (or averaging) and Floquet theory predictions, except at a
narrow stretch near the middle where both methods predict very different lines.
We tested this numerically and the Floquet theory results indeed indicate the
correct outcome.
On the red line a Hopf bifurcation occurs.
This is why the harmonic balance  approximation 
threshold predictions (see Appendix C) break down.
We believe this occurs because these approximations assume from the outset that
a period-doubling bifurcation occurs at the instability threshold.
We point out that if the dynamics of the coupled resonators could be decomposed into normal modes a Hopf bifurcation would not be possible. 
This is due to the constraints imposed on the multipliers by Liouville's formula.

\begin{figure}[!ht]
\centerline{\includegraphics[{scale=0.4}]{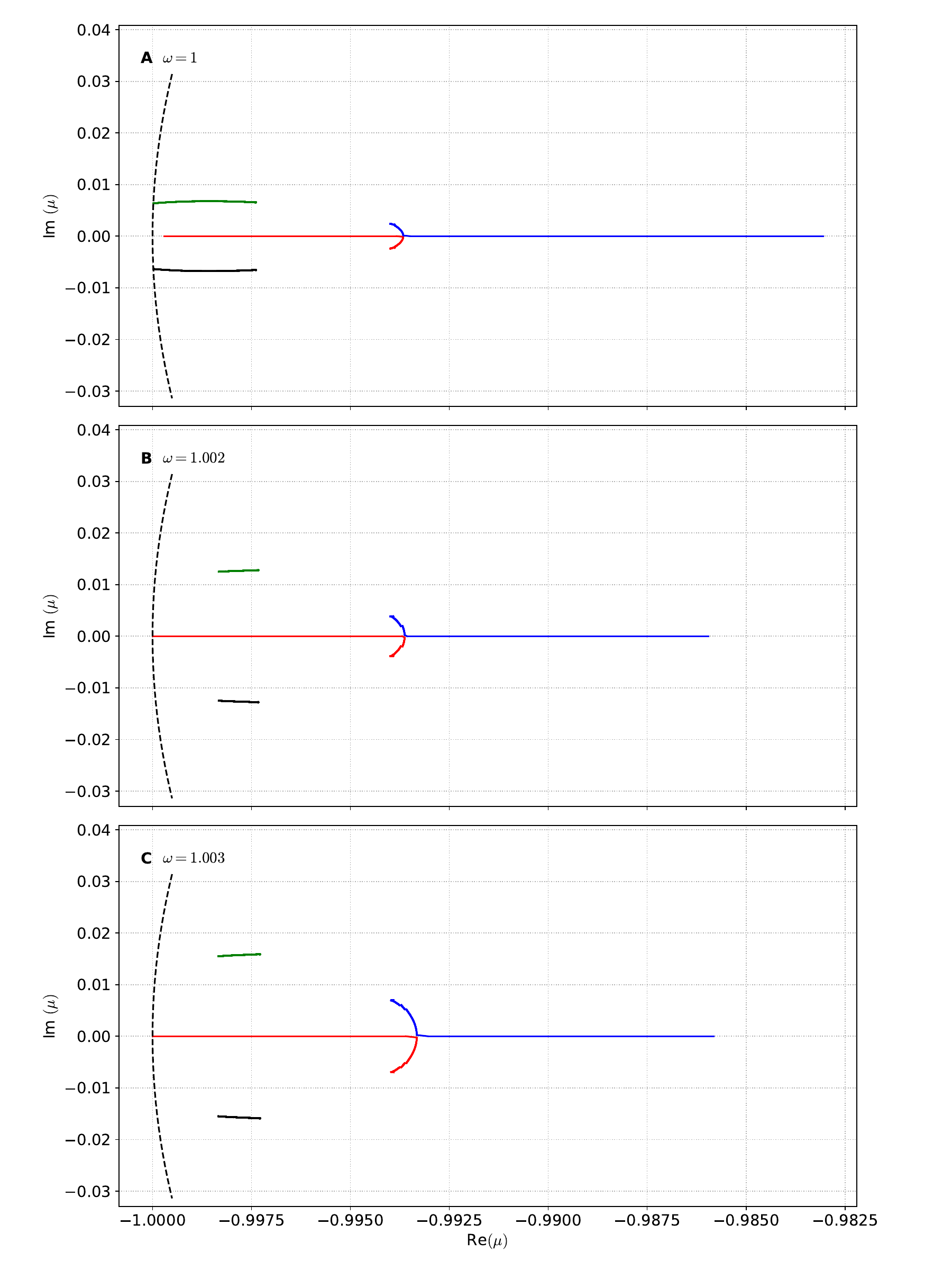}}
\caption{
Plots of the FMs in the complex plane as the pump amplitude is
varied up to the instability threshold. 
In panel {\bf A}, we have a Hopf bifurcation.
In panels {\bf B} and {\bf C}, we have a period doubling bifurcation.
The dashed line is an arc segment of the unit circle.
Here we used $T=\pi/\omega$.
}
\label{fig:hopf}
\end{figure}

In Fig. \ref{fig:hopf} we plot the FMs in the complex plane
as the pump amplitude $F_p$ is increased until one reaches the instability
threshold.
In panel {\bf A} a pair of complex conjugate FMs reaches
the unit circle while the real FMs have smaller amplitudes.
This is a hallmark of a Hopf bifurcation.
In panels {\bf B} and {\bf C} a real FM reaches -1 while the complex 
conjugate pair of FMs has amplitude less than one.
This characterizes a period-doubling bifurcation.

\begin{figure}[!ht]
\centerline{\includegraphics[{scale=0.6}]{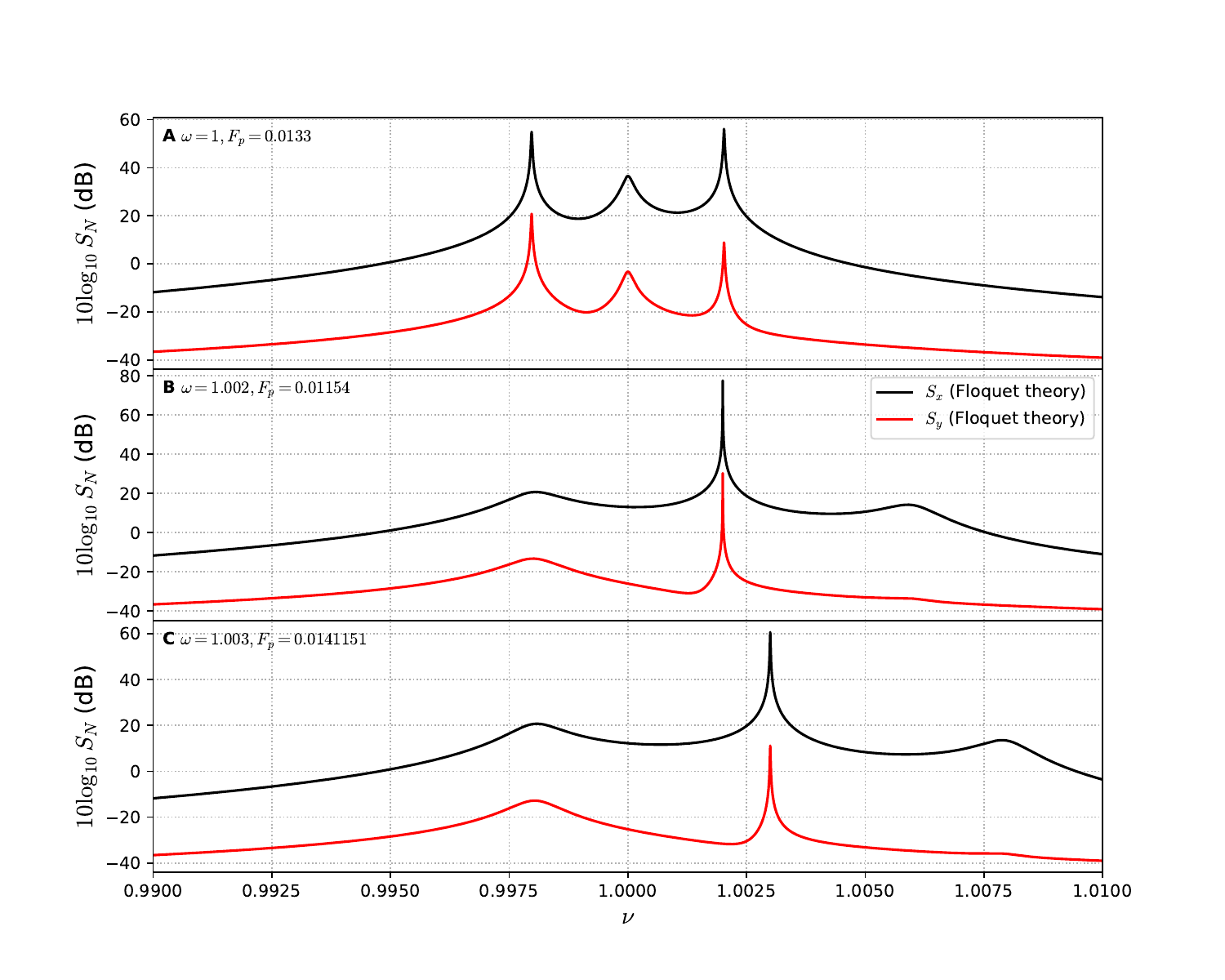}}
\caption{
Noise spectral densities for the parametric resonator ($S_x$) and the harmonic
resonator ($S_y$) of the coupled parametric-harmonic resonator model described
in Eqs. \eqref{2ModeParamp}.
These results are obtained very near the instability threshold.
In panel  {\bf A},  the left and right peaks are due to the complex 
conjugate pair of Floquet exponents and while the central peak at $\nu=1$ are
due to the two real Floquet exponents.
Since we have a Hopf bifurcation, the outer peaks dominate the spectrum while
the central peak is low and broad (as the real Floquet exponents are not zero).
In panels  {\bf B} and {\bf C}, one real Floquet exponent becomes zero at the
threshold to instability causing the central peak to become very narrow and
diverge.
Notice that in the Green's function method we have to use $T=2\pi/\omega$.   
}
\label{fig:NSD}
\end{figure}

In Fig. \ref{fig:NSD} we plot the spectrum of the NSDs of the position
coordinates of both resonators.
The three peaks indicate that two of the FMs are complex and two are real.
In all cases the instability threshold is very near in parameter space.
In panel {\bf A} one is close to a Hopf bifurcation.
This is the reason why the two peaks at $\omega\pm \Im\rho_c$ dominate the
spectrum, where $\rho_c$ and $\rho_c^*$ are the complex conjugate pair of FMs.
Although these peaks are large they do not diverge at the threshold because
their real part is not zero.
These results can be inferred from Eqs. \eqref{ABcoefs}.
Please see Ref.~\cite{batista2024amplification} for more details.
In panels {\bf B} and {\bf C} one is near a period-doubling bifurcation.
In this case the dominant peak becomes the one at $\nu=\omega$ and they
diverge at the threshold.
These peaks should be even higher than depicted. 
This is because they are extremely narrow and the sampling of the spectrum is
discretized.

\begin{figure}[!ht]
\centerline{\includegraphics[{scale=0.4}]{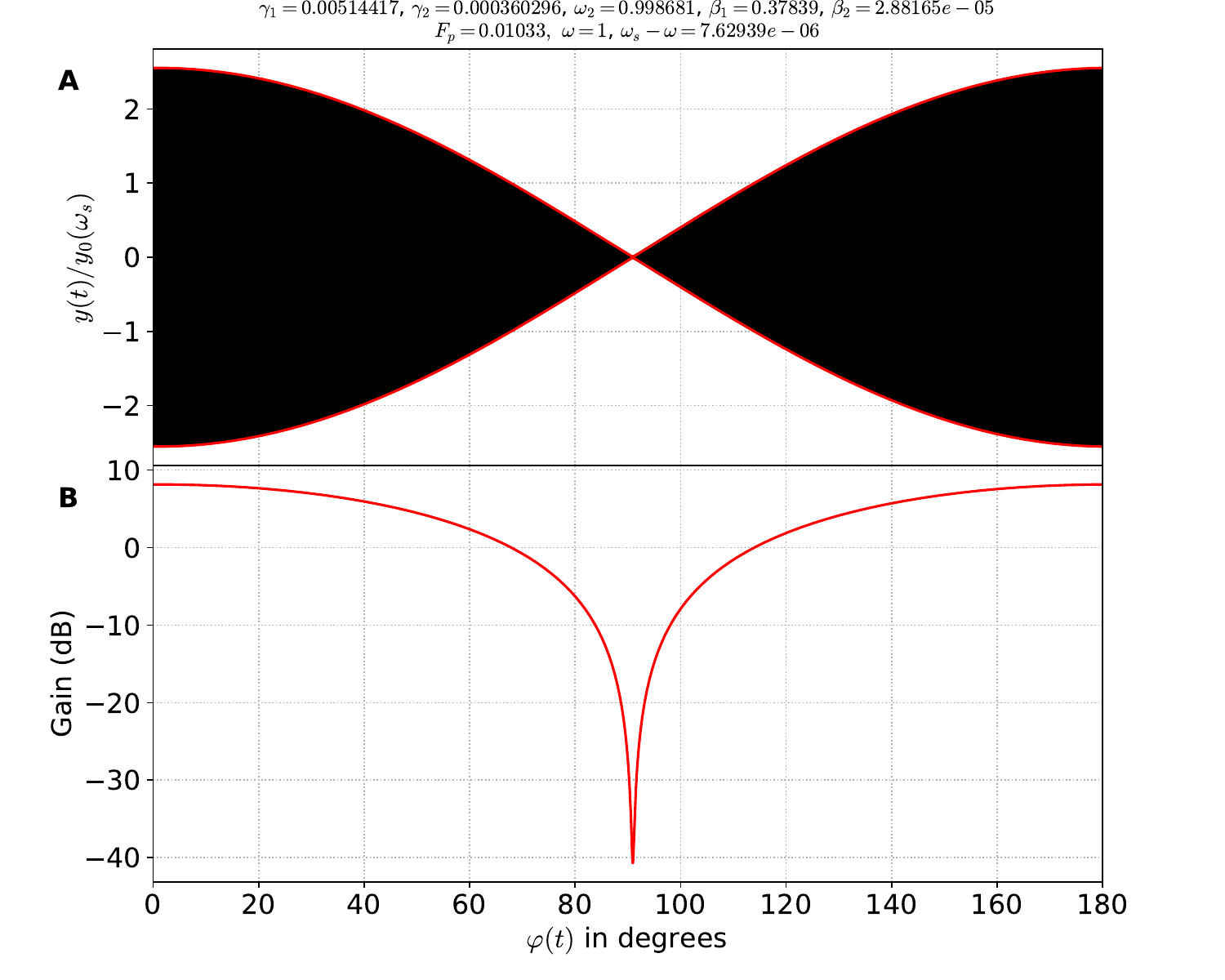}}
\caption{Gain as a function of phase for the parametrically modulated coupled
resonators described in Eq.~\eqref{2ModeParamp}, with $r_1(t)=0$ and $r_2(t)$
replaced by $\cos(\omega t+\varphi(t))$. 
{\bf A} We plot the normalized cyclo-stationary response $y(t)/y_0$ as a
function of phase $\varphi$, where $y_0$ is the amplitude of oscillations of
the forced harmonic resonator when $F_p=0$.
There is a slight detuning between $\omega$ and $\omega_s$ so that the phase is
swept very slowly, almost quasi-statically.
The envelopes are obtained from Floquet theory.  
{\bf B} We use the positive envelope from {\bf A} to obtain the gain in
decibels.
}
\label{fig:gainPhase}
\end{figure}

In Fig. \ref{fig:gainPhase} we show the phase sensitivity of the amplifier
based on the coupled parametric-harmonic rasonator model of
Eq.~\eqref{2ModeParamp}.
Here, $r_1(t)=0$ and $r_2(t)$ are replaced by $\cos(\omega t+\varphi(t))$.
This is similar to the method used by Rugar and Grütter to determine squeezing
from the coherent response dependent on phase in degenerate parametric
amplification.
We use this approach because stochastic simulations are expensive in terms of
computational resources and time.
We note that this is an accurate verification of the squeezing model for the coupled
resonators only if there is only one added noise input and if the detuning is
very small. 
Here, $\dot\varphi=\omega_s-\omega=2^{-17}\omega$, in such
a way that the phase is swept almost quasi-statically, what emulates the
method of Rugar and Grütter in only one integration run after the transients
die out.
In panel {\bf A}, we show a long time series of $y(t)$ divided by the constant
amplitude $y_0$ of $y(t)$ when $F_p=0$. 
In panel {\bf B}, we use the envelope of $y(t)/y_0$ to obtain the gain
in decibel.
We obtain a deamplification around $-40~$dB, which is a much deeper squeezing
than the original results of Rugar and Grütter.
Unlike previous results with deep squeezing, we did not use any feedback.
When one externally drives only the parametric resonator, only modest squeezing above $-6~$dB is obtained.

\begin{figure}[!ht]
\centerline{\includegraphics[{scale=0.4}]{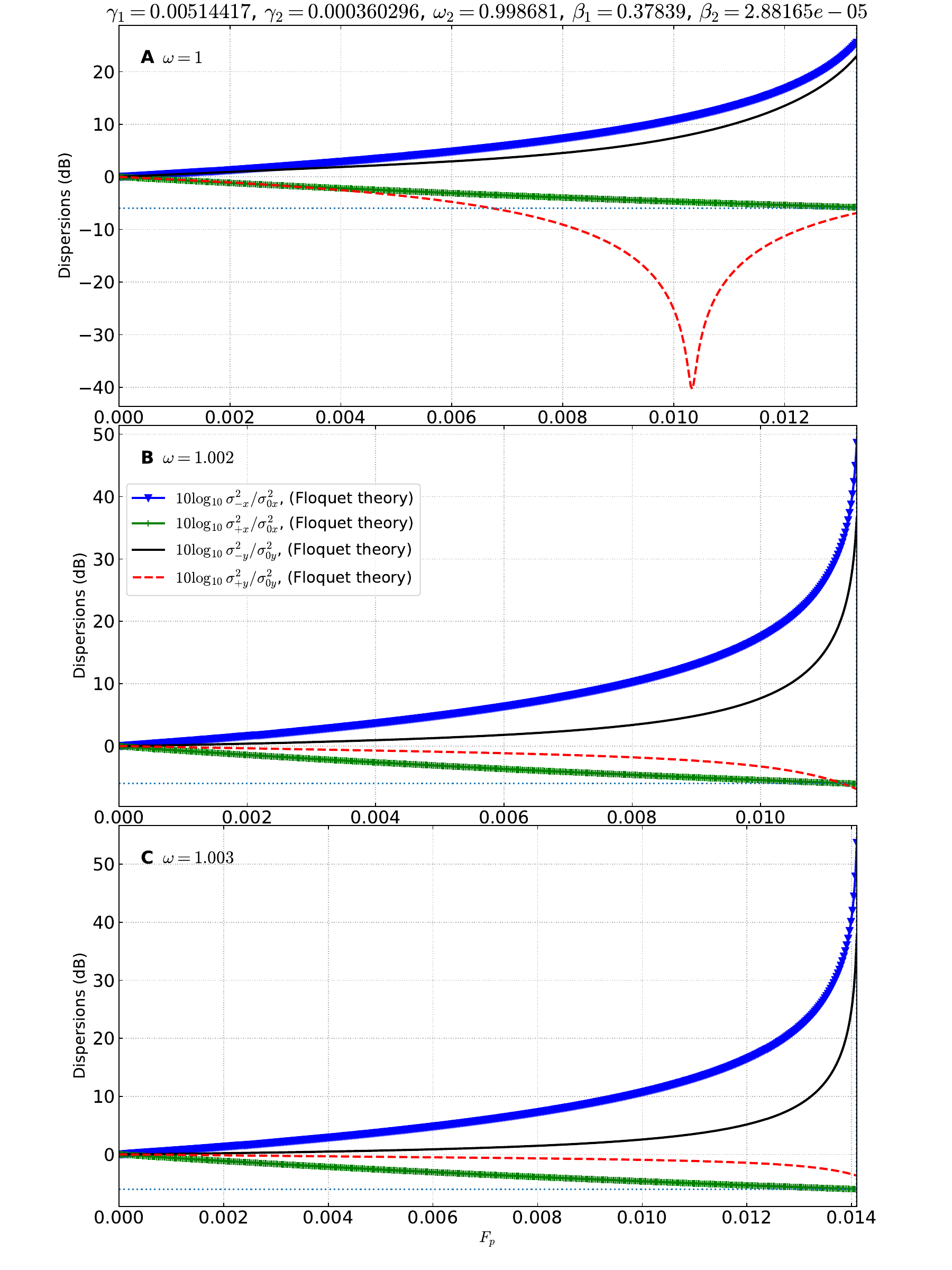}}
    \caption{
Floquet theory predictions for the model given in Eq.~\eqref{2ModeParamp}. 
Standard deviations in decibels relative to the equilibrium values with zero
pump in decibel scale.
The dotted horizontal line corresponds to $-6$~dB.
Each plot ends very close to the instability threshold. 
The common parameters used in all panels are given on top of the figure.
In panel (a), $\omega=1$, we see that, unlike the single-degree-of-freedom
parametric resonator, deep squeezing far below $-6~$dB can be achieved in the harmonic resonator.
In panel (b), $\omega=1.002$,  the
maximum squeezing at threshold (depicted by the vertical dashed line) is
slightly below $-6$~dB.
In panel (c), $\omega=1.003$, one sees that even with detuning one can reach
$-6$~dB squeezing limit near the instability threshold.
    }
\label{fig:si_uv2}
\end{figure}

In Fig. \ref{fig:si_uv2} we show the results of squeezing for the coupled
parametric-harmonic resonator model.
We can see in panel {\bf A}  that very deep sub-threshold squeezing.
This has the added advantage over previous results in the
literature that the deepest squeezing occurs somewhat before the onset of
oscillations.
Furthermore, the heating is bounded in all channels up to the frontier of
instability unlike what happens in the single parametric resonator.
The boundedness of heating is ultimately due to the fact that a Hopf
bifurcation occurs at the instability threshold for this choice of parameters
with $\omega=1$.
This feature is possible in coupled parametrically-driven resonator systems
whose dynamics cannot be simplified to a superposition of uncoupled normal
modes.
In this figure, we show results for the diagonalized dispersions which are
obtained from the dispersions for each one of the four channels (cosine and
sine channels of $x$ and $y$) and the correlations between the fluctuations of
the cosine and sine channels for each resonator.
The parameters used were taken from Singh \etal \cite{singh2018motion}, except
that we inverted the signal of the coupling parameters.
No attempt was made to make a thorough investigation of parameter space.
The theoretical model proposed here could be used, in principle, to quickly
search for parameters that provide the highest squeezing.
The utility of our model over brute numerical integration of stochastic
differential equations becomes more evident when one is near the instability
threshold, because at least one decay rate, the 
real part of a Floquet exponent, goes to zero.
This results in very long times for transients to die off, what becomes very
costly in terms of computational resources.

\begin{figure}[!ht]
\centerline{\includegraphics[{scale=0.5}]{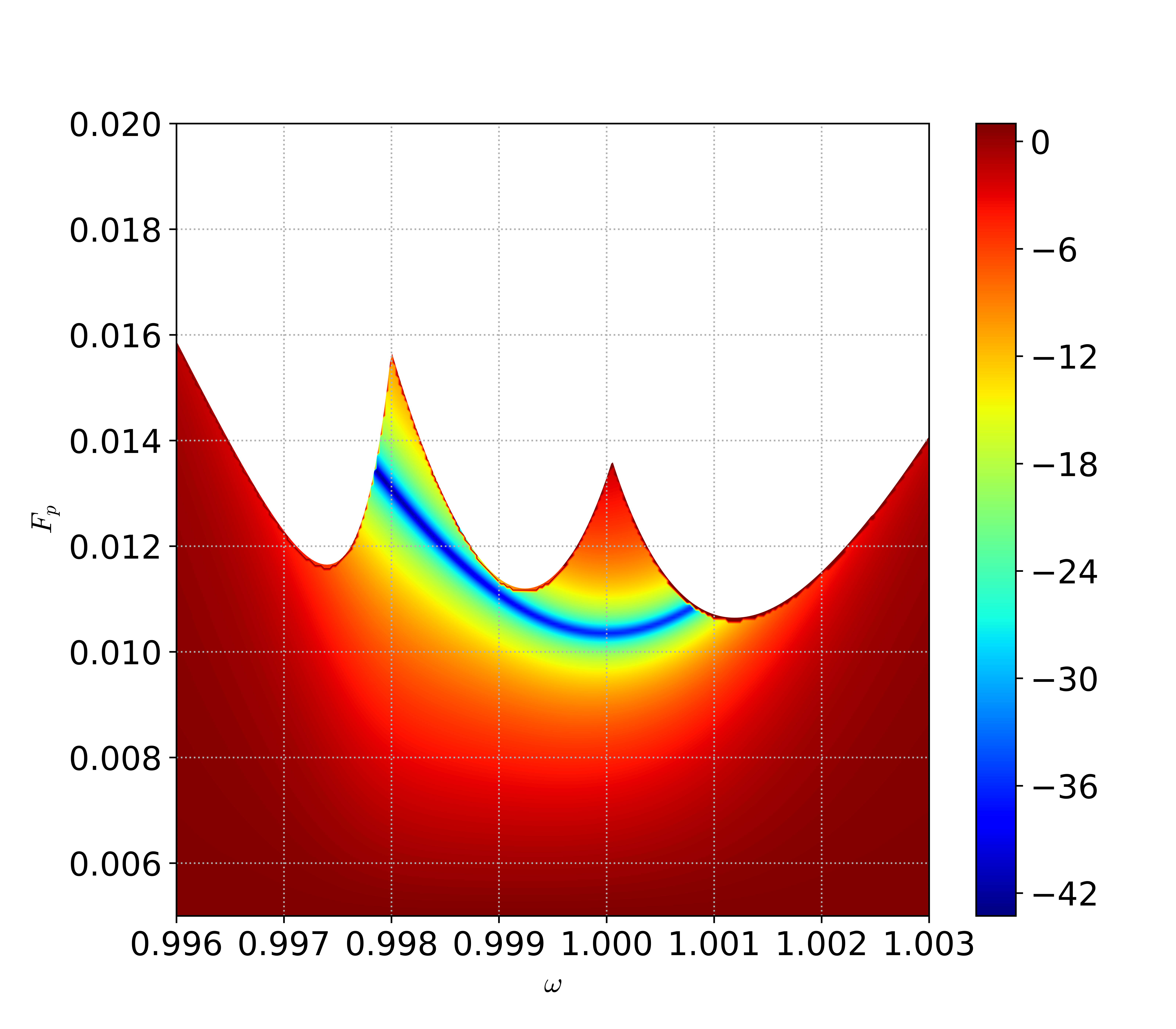}}
    \caption{Two dimensional color plot of the gain function 
$10\log_{10}\sigma_{+y}^2/\sigma^2_{0y}$ in the $(\omega, F_p)$ parameter space.
The color bar scale is in decibel.
The fixed parameters are the same as in the previous figure.
The threshold line to instability is a small portion of the one depicted in
Fig. \ref{fig:boundary}, near the central part where a Hopf bifurcation occurs.
}
\label{fig:squeezingMap}
\end{figure}

In Fig. \ref{fig:squeezingMap} we make a 2D color plot in $(\omega, F_p)$ 
parameter space to better visualize the region with deep squeezing.
We see that this is mostly below the stretch of the threshold line where
a Hopf bifurcation occurs.
This approach may be very helpful for the experimentalist to design an
experiment to find deep squeezing.


\FloatBarrier
\section{Conclusion}
\label{conclusion}
Here we developed a theoretical model based on stochastic differential equations
to explain the effect of white noise squeezing that
occurs in classical parametrically driven resonators with added noise.
This phenomenon, which is akin to the quantum phenomenon of squeezed coherent
states, was first observed in a classical micromechanical resonator in 1991 by
Rugar and Grütter \cite{rugar91}.
Unlike previous works, our results are based on the solution of stochastic
differential equations, approximately, using the averaging method and,
exactly, using Floquet theory.
Another novelty of our model is that we obtain analytical expressions for
statistical averages related to the frequency-domain response of the
parametrically modulated resonator to added white noise. 
Each data point corresponds to a Fourier transform of a long time series
with coordinates, one for each input signal in quadrature, given by the
real and imaginary parts of the Fourier transform of the parametric resonator
response to the added noise.
The experimental case approximates by the ideal theoretical case when
the time constant of the LIA goes to infinite.

For the case of a single-degree-of-freedom parametric resonator we obtained
excellent agreement between the predictions of the 1st-order averaging
approximations and Floquet theory near the first parametric instability when
$\omega\approx1$ (in our dimensionless units).
The strongest squeezing occurs at $\omega=1$ and $F_p$ set
at the threshold of parametric instability.
While in one quadrature, fluctuations are decreased as the pumping
amplitude increases, reaching a $-6$~dB deamplification in amplitude at the
parametric instability threshold, in the other quadrature, fluctuations can
increase without bounds.
We showed that, unlike previously implied in the literature, squeezing does
not quickly disappear with increasing (red or blue) detuning.
We reached this conclusion by taking into account the effect of correlation,
what to the best of our knowledge, had not been properly investigated in the
literature.
We believe this discovery could be relevant for the experimentalists.
It is important to point out that the noise squeezing effect increases when
one nears the parametric instability threshold, as can be seen in Eq.~\eqref{sigma2_s_om_approx}.
We showed here that this is a feature of amplification near a period-doubling bifurcation point.
It could be a generic feature of local codimension-1 bifurcation points
and could occur in physical systems such as the Josephson junction amplifier
\cite{yamamoto2008flux, vijay2009invited}.

We developed a general model of squeezing based on Green's
functions and Floquet theory that can be applied to any number of
coupled parametrically modulated resonators with added noise.
Here we obtained deep squeezing, even beyond the lower limit previously obtained
in parametric systems with feedback \cite{poot2015deep, patil2015thermomechanical, sonar2018strong}.
We showed that the strong squeezing we found is only possible in coupled
resonators whose dynamics cannot be decomposed into normal modes.
In those situations, one cannot obtain deamplification below $-6~$dB.
Furthermore, it seems that strong squeezing only occurs when at least two 
FMs are complex.
In addition to that, the squeezing seems to be strongest in regions close to
where the bifurcation to instability is a Hopf bifurcation and not a
period-doubling bifurcation.
In those situations, the fluctuations in all quadratures do not diverge
at the instability threshold.
We believe that our model can be applied to a wide number of
parametrically-driven resonators and could be used as a guide to
experimentalists in searching suitable parameter regions with deep squeezing.

The methods developed here could be applied to investigate quadrature squeezing
in limit cycles of nonlinear oscillators perturbed by added noise.
This could be implemented experimentally in nanomechanical resonators.
Also, more interestingly, since the quantum limit of these resonators has been
studied in the last 10-15 years \cite{oconnell2010quantum,samanta2023nonlinear},
one could investigate what the quadrature squeezing would be in those systems.
Another system of interest is the Kerr parametric oscillator which has been
implemented in superconducting circuits based on Josephson junctions
\cite{goto2019quantum,grimm2020stabilization}.
This system is the quantum equivalent of a parametrically pumped Duffing
oscillator \cite{batista2020frequency}.
Since networks of coupled parametrically-driven resonators (perceptrons) have
been proposed as a way to implement Ising machines \cite{heugel2022ising}, the
accurate knowledge of the threshold boundary has become a more relevant issue.
The Floquet theory method developed here could be used to obtain the 
instability line accurately and efficiently, and also be used to the predict
the amount of noise squeezing involved.
Also, it could be used in conjunction with feedback to achieve even stronger
squeezing \cite{mashaal2024strong}.

\section*{Supplementary Material}
The code used to generate all the numerical data and the figures of this paper 
will be provided in the supplementary material.
\section*{Appendix}
\setcounter{equation}{0}
\renewcommand\theequation{A\arabic{equation}}
\subsection{The Gaussian distribution with two variables}
\label{gaussian2var}
Given the Gaussian bivariate probability distribution of the type
\beq
P(x, y)= \frac{\sqrt{ab-c^2}}{\pi} \,e^{-ax^2-by^2-2cxy},
\label{eq:P_x1x2}
\eeq
where $a,b>0$ and $ab>c^2$, we find $\langle x\rangle=\langle y\rangle=0$ and
\beq
\begin{aligned}
\langle x^2\rangle  &= \frac{b}{2(ab-c^2)},\\
\langle y^2\rangle  &= \frac{a}{2(ab-c^2)},\\
\langle xy\rangle &=-\frac{c}{2(ab-c^2)}.
\end{aligned}
\eeq

Inverting these equations, we obtain all the parameters of this distribution
\beq
\begin{aligned}
a &= \frac{\langle y^2\rangle}{2\left(\langle x^2\rangle\langle y^2\rangle-\langle xy\rangle^2\right)},\\
b &= \frac{\langle x^2\rangle}{2\left(\langle x^2\rangle\langle y^2\rangle-\langle xy\rangle^2\right)},\\
c &= -\frac{\langle xy\rangle}{2\left(\langle x^2\rangle\langle y^2\rangle-\langle xy\rangle^2\right)}.
\end{aligned}
\eeq
Consequently, once we have the experimental data we can find the statistical
averages $\langle x^2\rangle$, $\langle y^2\rangle$, and $\langle xy\rangle$, and from there obtain the probability distribution.
We can also diagonalize the bilinear form at the exponent of the probability
distribution.
In other words we want to split the joint probability distribution into the
product of two one variate distributions, that is $P(x,y)=P(u, v)=P_-(u)P_+(v)$
such that $ax^2+by^2+2cxy=\lambda_-u^2+\lambda_+v^2$.
To achieve that, we  diagonalize the following matrix
\beq
A=
\left[
\bea{cc}
    a &c\\
    c &b
\ea
\right].
\eeq
The characteristic equation is $\lambda^2-\Tr A\lambda+\det A=0$, from where
we obtain the eigenvalues
\beq
\begin{aligned}
\lambda_\pm&=\frac{\Tr A\pm\sqrt{\Tr A^2-4\det A}}2
=\frac{a+b\pm\sqrt{(a-b)^2+4c^2}}2\\
&=\frac{\langle x^2\rangle+\langle y^2\rangle\pm\sqrt{\left(\langle x^2\rangle-\langle
y^2\rangle\right)^2+4\langle xy\rangle^2}}{4\left(\langle x^2\rangle\langle
y^2\rangle-\langle xy\rangle^2\right)}.
\end{aligned}
\label{la_pm}
\eeq
The new distribution is
\beq
P(u, v)=\frac{\sqrt{\lambda_-\lambda_+}}\pi e^{-\lambda_-u^2-\lambda_+v^2}
\eeq
and the new dispersions are
\beq
\begin{aligned}
\sigma_-^2=\langle u^2\rangle=\frac1{2\lambda_-},\\
\sigma_+^2=\langle v^2\rangle=\frac1{2\lambda_+}.
\end{aligned}
\label{eq:dispersions}
\eeq
The normalized eigenvectors are given by
\beq
v_\pm=\frac1{\sqrt{1+(\lambda_\pm-a)^2/c^2}}
\left(
\bea{c}
1\\
\frac{\lambda_\pm-a}c
\ea
\right).
\label{v_mv_p}
\eeq
\beq
D=\left(
\bea{cc}
\lambda_-&0\\
0&\lambda_+
\ea
\right)=S^TAS,
\eeq
where
\beq
S=\Bigl[v_-\;\vline\; v_+\Bigr].
\eeq
We then have
\beq
\renewcommand\arraystretch{0.8}
\begin{bmatrix}
u\\v
\end{bmatrix}
=
\renewcommand\arraystretch{0.6}
\begin{bmatrix}
v_-^T\\
---\\
v_+^T
\end{bmatrix}
\renewcommand\arraystretch{0.8}
\begin{bmatrix}
x\\y
\end{bmatrix}
\eeq
We can also assert that the tangent of the angle of the eigenvectors with
respect to the $x$ axis is given by
\beq
\tan\theta_\pm=\frac{\lambda_\pm-a}c= \frac{b-a\pm\sqrt{(b-a)^2+4c^2}}{2c}=\frac{\langle y^2\rangle-\langle x^2\rangle\mp\sqrt{\left(\langle x^2\rangle-\langle
y^2\rangle\right)^2+4\langle xy\rangle^2}}{2\langle xy\rangle}
.
\eeq
We notice that when $a=b$, $\theta_\pm=\pm\pi/4$.
\subsection{The Gaussian distribution with $N$ variables}
Given the Gaussian $N$-variate probability distribution of the type
\beq
P(x_1, x_2,\dots, x_N)= \frac1{\sqrt{(2\pi)^N\det C}}\,e^{-\frac12\sum_{jk}C^{-1}_{jk}x_jx_k},
\label{eq:P_N}
\eeq
where $A$ is a positive-defined matrix (i.e. where all eigenvalues are positive).
The dispersions are given by
\beq
\begin{aligned}
\langle x_ix_j\rangle&=
\frac1{\sqrt{(2\pi)^N\det C}}\int_{-\infty}^\infty\dots\int_{-\infty}^\infty
x_ix_j\,e^{-\frac12\sum_{jk}C^{-1}_{jk}x_jx_k}dx_1\dots dx_N\\
&=\frac1{\sqrt{(2\pi)^N\det C}}\frac{\partial^2}{\partial\lambda_i\partial\lambda_j}
\left.\int_{-\infty}^\infty\dots\int_{-\infty}^\infty
\,e^{-\frac12\sum_{jk}C^{-1}_{jk}x_jx_k+\sum_k\lambda_kx_k}dx_1\dots
dx_N\right|_{\lambda_1=...=\lambda_N=0}\\
&=\left.\frac{\partial^2}{\partial\lambda_i\partial\lambda_j}e^{\frac12\sum_{jk}C_{jk}\lambda_j\lambda_k}\right|_{\lambda_1=...=\lambda_N=0}=C_{ij}
\end{aligned}
\eeq
In order to obtain this result, we made the following change of coordinates
$x_j=y_j+\alpha_j$.
Thus, the exponent of the distribution from \eqref{eq:P_N} becomes
\[
\begin{aligned}
S&=-\frac12\sum_{jk}C^{-1}_{jk}x_jx_k+\sum_k\lambda_kx_k
=-\frac12\sum_{jk}C^{-1}_{jk}(y_j+\alpha_j)(y_k+\alpha_k)+\sum_k\lambda_k(y_k+\alpha_k)\\
&=-\frac12\sum_{jk}C^{-1}_{jk}y_jy_k-\sum_{j}\left[\sum_kC^{-1}_{jk}\alpha_k-\lambda_j\right]y_j-\frac12\sum_{jk}C^{-1}_{jk}\alpha_j\alpha_k+\sum_k\lambda_k\alpha_k.
\end{aligned}
\]
The $\alpha_j$ are chosen such that 
\[
\sum_kC^{-1}_{jk}\alpha_k-\lambda_j=0
\implies 
\bm{\alpha= C\lambda}
\]
We obtain then
\[
S=-\frac12\sum_{jk}C^{-1}_{jk}y_jy_k+\frac12\sum_k\lambda_k\alpha_k
=-\frac12\sum_{jk}C^{-1}_{jk}y_jy_k+\frac12\sum_{jk}C_{jk}\lambda_k\lambda_k.
\]
\subsection{The harmonic balance approximation to the the threshold line}
At the threshold line of the coupled resonator model, whose dynamics is
governed by Eqs. \eqref{2ModeParamp} with $r_1(t)=r_2(t)=0$, we seek stationary solutions given by 
\beq
\begin{aligned}
x(t) &= \frac12\left(A_1e^{i\omega t}+A_1^*e^{-i\omega t}\right),\\
y(t) &= \frac12\left(A_2e^{i\omega t}+A_2^*e^{-i\omega t}\right).
\end{aligned}
\eeq
The coefficients of these stationary solutions obey
\beq
\begin{aligned}
    &\left(1-\omega^2+i\gamma_1\omega\right)A_1-\beta_1A_2+i\frac{F_p}2A_1^*=0,\\
&\left(\omega_2^2-\omega^2+i\gamma_2\omega\right)A_2-\beta_2A_1=0.\\
\end{aligned}
\eeq
which implies into
\beq
\begin{aligned}
    &\alpha(\omega)A_1+i\frac{F_p}2A_1^*=0,\\
    &-i\frac{F_p}2A_1+\alpha^*(\omega)A_1^*=0,
\label{system1}
\end{aligned}
\eeq
where
\beq
\alpha(\omega)=1-\omega^2+i\gamma_1\omega-\frac{\beta_1\beta_2}{\left(\omega_2^2-\omega^2+i\gamma_2\omega\right)}.
\eeq
At the threshold line we find
\beq
|\alpha(\omega)|^2-F_p^2/4=0.
\label{har_bal_threshold}
\eeq
\FloatBarrier

%
\end{document}